\documentclass[12pt]{JHEP3}
\title{Solar Neutrinos Before and After Neutrino 2004}

\author{John N. Bahcall\\
  School of Natural Sciences, Institute for Advanced Study, Princeton,
  NJ 08540\\
    E-mail: \email{jnb@ias.edu}}
\author{M. C. Gonzalez-Garcia \\
  C.N. Yang Institute for Theoretical Physics\\
State University of New York at Stony Brook\\
Stony Brook,NY 11794-3840, USA,\\
and Instituto de F\'\i sica Corpuscular, Universitat de Val\`encia
  -- C.S.I.C. \\ Edificio Institutos de Paterna, Apt 22085, 46071
  Val\`encia, Spain\\
    E-mail: \email{concha@insti.physics.sunysb.edu}}
\author{Carlos Pe\~na-Garay\\
        School of Natural Sciences, Institute for Advanced Study,
  Princeton, NJ 08540\\
       E-mail: \email{penya@ias.edu}}

\abstract{We compare, using a three neutrino analysis, the allowed
neutrino oscillation parameters and solar neutrino fluxes
determined by the experimental data available  Before and After
Neutrino~2004. New data available after Neutrino~2004 include
refined KamLAND and gallium measurements. We use six different
approaches to analyzing the KamLAND data. We present detailed
results using all the available neutrino and anti-neutrino data
for $\Delta m^2_{21}$, $\tan^2 \theta_{12}$, $\sin^2 \theta_{13}$,
and $\sin^2 \eta$ (sterile fraction). Using the same complete data
sets, we also present Before and After determinations of all the
solar neutrino fluxes  (which are treated as free parameters), an
upper limit to the luminosity fraction associated with CNO
neutrinos, and the predicted rate for a $^7$Be solar neutrino
experiment. The $1\sigma$ ($3\sigma$) allowed range of $\Delta
m^2_{21} = 8.2^{+0.3}_{-0.3}(^{+{1.0}}_{{-0.8}}) \times 10^{-5} \,
{\rm eV^2}$ is decreased by a factor of 1.7 (5), but the allowed
ranges of all other neutrino oscillation parameters and neutrino
fluxes are not significantly changed. Maximal $\theta_{12}$ mixing
is disfavored  at 5.8$\sigma$ and the bound on the mixing angle
$\theta_{13}$ is slightly
    improved to $\sin^2\theta_{13} <0.048$ at 3$\sigma$. The predicted rate
in a $^7$Be neutrino-electron scattering experiment is $0.665\pm
0.015\, (^{+0.045}_{-0.040})$ of the rate implied by the BP04
solar model in the absence of neutrino oscillations. The
corresponding predictions for $p-p$ and $pep$ experiments are,
respectively,   $0.707 ^{+0.011}_{-0.013} (^{+0.041}_{-0.039})$
and  $0.644 ^{+0.011}_{-0.013} (^{+0.045}_{-0.037})$. In order to
clarify what measurements constrain which parameters best, we also
analyze the solar neutrino data separately and the reactor
anti-neutrino data separately, both Before and After
Neutrino~2004. We derive upper limits to CPT violation in the weak
sector by comparing reactor anti-neutrino oscillation parameters
with neutrino oscillation parameters. We also show that the recent
data disfavor at 91\% CL a  proposed non-standard interaction
description of solar neutrino oscillations. We have verified that
our results are insensitive (changes much less than $1\sigma$) to
which of six approaches  we use in analyzing the KamLAND data,
which of the published $^8$B neutrino energy spectra we adopt, and
the precise value of the gallium solar neutrino event rate.}


\begin{document}
\input psfig
\widowpenalty=200

\section{Introduction}
\label{introduction}

How have the recently released data from the  the KamLAND reactor
anti-neutrino experiment~\cite{newkamland}  and the revised
average gallium solar neutrino
rate~\cite{sage04,sage02,gallex,gno} improved our knowledge of
neutrino properties and of solar neutrino fluxes?

We concentrate in this paper on a Before-After comparison that is
made possible by the new data released at Neutrino 2004 (Paris,
June 19--24, 2004). We determine how the new KamLAND and solar
neutrino data affect our knowledge of the parameters that
characterize solar neutrino oscillations [$\Delta m^2_{21}$,
$\theta_{12}$, $\theta_{13}$, $\eta_{\rm sterile}$] and the
parameters that characterize solar energy generation and neutrino
fluxes [$L_{\rm CNO}$, $\phi(p-p)$, $\phi(^7{\rm Be})$,
$\phi(^8{\rm B})$, $\phi(^{13}{\rm N})$,  $\phi(^{15}{\rm O})$,
and  $\phi(^{17}{\rm F})$].

In order to clarify which measurements constrain what quantities
and by how much, we analyze the reactor
data~\cite{newkamland,kamlandfirstpaper,chooz} separately and  the
solar neutrino
data~\cite{sage04,sage02,gallex,gno,chlorine,kamiokande,superk,snofirstpaper,snosalt,snoccnc,snodaynight}
separately. We use six different approaches  to analyzing the
KamLAND data (see  section~\ref{subsec:fivekamland}) in order to
assess the quantitative importance, or lack of importance, of
different analysis procedures.

The conventional wisdom is that a quantitative improvement in our
knowledge of neutrino parameters and solar neutrino fluxes is all
that we should expect. According to this view, the existing solar
neutrino experiments have reached a level of maturity and
precision at which new data from these operating experiments are
expected to lead to refinements, but not revolutions. This
conventional wisdom could be wrong and sub-dominant contributions
due, for example, to non-standard
interactions~\cite{nsi,othernsi}, to sterile
neutrinos~\cite{smirnovsterile,sterileberezinsky}, or even to CPT
violation~\cite{bahcall:2002ia} could show up in the operating
experiments. We investigate all these possibilities.

We analyze the experimental data assuming that vacuum and matter
neutrino oscillations~\cite{pontecorvo,msw} occur among three
active neutrino species (with the possibility also of oscillation
to a sterile neutrino~\cite{four}). The techniques that we use in
this analysis have been described previously in a series of
papers, especially refs.~\cite{postkamland,cnopaper,roadmap}. Many
other groups have reported analyses of solar neutrino and reactor
data, see ref.~\cite{otherspostsnoII}, but our analysis is unique
so far-we believe-in treating all of the solar neutrino fluxes as
free parameters, subject only to the luminosity constraint,
refs.~\cite{luminosity,spirovignaud} (i.e., essentially energy
conservation).

Our principal results are shown in figure~\ref{fig:solar_kamland}
and figure~\ref{fig:SplusK} and  in table~\ref{tab:globalosc} and
table~\ref{tab:globalfluxes}.

We begin by discussing in section~\ref{sec:experimentalandchi2}
the experimental data and the $\chi^2$ formulations we use for
different applications. The data are summarized in
table~\ref{tab:experimental}. We then present in
section~\ref{sec:newkamland} the results of our reactor-only
analyses: the allowed regions in neutrino oscillation space that
are compatible with the reactor data available Before
Neutrino~2004 and the reactor data (notably the new KamLAND
data~\cite{newkamland}) available After Neutrino~2004. We also
describe in section~\ref{sec:newkamland} the six different
approaches we use in analyzing the KamLAND data and summarize the
technical aspects of our KamLAND analyses. Next we present in
section~\ref{sec:newsolar} the results of our solar-only analyses:
the allowed regions in neutrino parameter space and neutrino
fluxes that are compatible with all solar neutrino data available
both Before and After Neutrino~2004.
Figure~\ref{fig:solar_kamland} summarizes the results of our
Before-After reactor-only and solar-only comparisons.

We present in section~\ref{sec:newglobal} and in
table~\ref{tab:globalosc} and table~\ref{tab:globalfluxes} the
results of our global three neutrino analyses of solar neutrino
experimental results and reactor anti-neutrino data. We give the
best-estimates and the uncertainties for neutrino oscillation
parameters and for solar neutrino fluxes. We also determine in
this section the upper bound on the sterile neutrino flux and on
the luminosity of the Sun that is associated with the CNO nuclear
fusion reactions. We compare in section~\ref{sec:cpt} the allowed
oscillation regions of rector anti-neutrinos  with the allowed
oscillation regions of solar neutrinos in order to establish an
upper limit on CPT violation in the weak sector. We summarize and
discuss our main results in section~\ref{sec:summary}.  In the
Appendix, we present some details of the analysis involving
$\theta_{13}$.

\section{Experimental data and $\chi^2$}
\label{sec:experimentalandchi2}

\TABLE[!t]{
 \caption[]{{\bf Experimental data.} We summarize the solar,
reactor, accelerator, and atmospheric data used in our global
analyses. Only experimental errors are included in the column
labelled Result/SM. Here the notation {SM} corresponds to
predictions of the Bahcall-Pinsonneault standard solar model
(BP04) of ref.~\protect\cite{bp04}  and the standard model of
electroweak interactions~\cite{sm} (with no neutrino
oscillations). The new average gallium rate is $68.1\pm3.75$ SNU
(see ref.~\cite{sage04}). The SNO rates (pure $D_2O$ phase) in the
column labelled Result/SM are obtained from the published SNO
spectral data by assuming that the shape of the ${\rm ^8B}$
neutrino spectrum is not affected by physics beyond the standard
electroweak model. However, in our global analyses, we allow for
spectral distortion. The SNO rates (salt phase) are not
constrained to the $^8$B shape \protect\cite{snosalt}. The K2K and
atmospheric data are used only in the analysis of $\theta_{13}$,
which is discussed in Appendix~\ref{sec:theta13}.
\label{tab:experimental}
\\}\footnotesize
\begin{tabular}{llll}
\hline\noalign{\smallskip}
\multicolumn{1}{c}{Experiment} &\multicolumn{1}{c}{Observable ($\#$
  Data)} &\multicolumn{1}{c}{Measured/SM} &\multicolumn{1}{c}{Reference}\\
\noalign{\smallskip}\hline\noalign{\smallskip}
 Chlorine & Average Rate (1) & [CC]=$0.30 \pm 0.03$ &\ \ \protect\cite{chlorine}\\
SAGE+GALLEX/GNO$^{\dagger}$ & Average Rate (1)& [CC]=$0.52 \pm 0.03$ &\ \ \protect\cite{sage02,gallex,gno}\\
Super-Kamiokande & Zenith Spectrum (44)& [ES]=$0.406 \pm 0.013$ &\ \ \protect\cite{superk}\\
SNO (pure {$ \rm D_2$O} phase) & Day-night Spectrum (34)&[CC]=$0.31 \pm 0.02$&\ \ \protect\cite{snoccnc,snodaynight}\\
 & &[ES]=$0.47 \pm 0.05$&\ \ \cite{snoccnc,snodaynight}\\
 & &[NC]=$1.01 \pm 0.13$&\ \ \cite{snoccnc,snodaynight}\\
SNO (salt phase) & Average Rates (3)&[CC]=$0.28 \pm 0.02$&\ \ \protect\cite{snosalt}\\
 & &[ES]=$0.38 \pm 0.05$&\ \ \cite{snosalt}\\
 & &[NC]=$0.90 \pm 0.08$&\ \ \cite{snosalt}\\
KamLAND & Spectrum (10)&[CC]=$0.69 \pm 0.06$&\ \ \protect\cite{kamlandfirstpaper}\\
CHOOZ & Spectrum (14) &{[CC] = $1.01 \pm 0.04$} & \ \ \protect\cite{chooz}\\
K2K& Spectrum (6) &[CC]$(\nu_\mu)$ = $0.70^{+0.11}_{-0.10}$ & \ \ \protect\cite{k2k}\\
Atmospheric & Zenith Angle Distributions (55) & [0.5-1.0] & \ \ \protect\cite{atm}\\
 \noalign{\smallskip}\hline\noalign{\smallskip}
\end{tabular}
\hbox to\hsize{$^{\dagger}$ SAGE rate: $66.9 \pm 3.9 \pm
3.6$~SNU~\protect\cite{sage02}; GALLEX/GNO rate: $69.3 \pm 4.1 \pm
3.6$~SNU~\protect\cite{gallex,gno}.  \hfill}
}

We summarize in this section the experimental data we use and the
$\chi^2$ distributions that we analyze.

Table~\ref{tab:experimental} summarizes the
solar~\cite{sage04,sage02,gallex,gno,chlorine,snoccnc,snodaynight},
reactor~\cite{newkamland,kamlandfirstpaper,chooz}, and
atmospheric~\cite{atm} data used in our global analyses that are
presented in section~\ref{sec:newglobal}. The number of data
derived from each experiment are listed (in parentheses) in the
second column of the table.  In the third column, labelled
Measured/SM, we list for each experiment the quantity Measured/SM,
the measured total rate divided by the rate that is expected
assuming the correctness of, as relevant, the standard solar model
and the standard model of electroweak interactions (i.e., no
neutrino oscillations or other non-standard physics).

We calculate the global $\chi^2$ by fitting to all the available
data, solar plus reactor. For the analysis of the upper bound on
$\theta_{13}$, we also include data from the K2K accelerator
experiment and from atmospheric measurements, see
eq.~(\ref{eq:chiglomarg}). Formally, the global $\chi^2$ can be
written in the form~\cite{chi2a,chi2b}

\begin{eqnarray}
\chi^2_{\rm global} &=& \chi^2_{\rm solar}(\Delta
m^2_{21},\theta_{12},\theta_{13}, \{ f_{\rm B}, f_{\rm Be},
f_{p-p}, f_{\rm CNO}\}) \nonumber\\
 &+& \chi^2_{\rm KamLAND}(\Delta m^2_{21},\theta_{12},\theta_{13}) \,.
\label{eq:chisquaredpowerful}
\end{eqnarray}

Depending upon the case we consider, there can be as many as nine
free parameters in $\chi^2_{\rm solar}$, including, $\Delta m^2_{21},
\theta_{12}$, $\theta_{13}$ , $f_{\rm B},$ $f_{\rm Be}$,
$f_{p-p}$, and $f_{\rm CNO}$ (3 CNO fluxes, see below). The
neutrino oscillation parameters $\Delta m^2_{21},
\theta_{12},\theta_{13}$ have their usual meaning. The reduced
fluxes $f_{\rm B}$, $f_{\rm Be}$, $f_{p-p}$, and $f_{\rm CNO}$ are
defined as the true solar neutrino fluxes divided by the
corresponding values of the fluxes predicted by the BP04 standard
solar model~\cite{bp04}. We extend in section~\ref{subsec:sterile}
the formalism to include sterile neutrinos.

The function $\chi^2_{\rm KamLAND}$ depends only on $\Delta
m^2_{21}$, $\theta_{12}$ and $\theta_{13}$.

We marginalize $\chi^2_{\rm global}$ making use of the function
$\chi^2_{\rm CHOOZ+ATM + K2K}(\theta_{13})$  that was obtained
following the analysis of ref.~\cite{3nuupdate} of
atmospheric~\cite{atm}, K2K accelerator~\cite{k2k}, and CHOOZ
reactor~\cite{chooz} data (see also,
refs.~\cite{noon04,ournewatm}).
 We have not assumed, as is often done, a
flat probability distribution for all values of $\theta_{13}$
below the CHOOZ bound. The fact that we take account of the actual
experimental constraints on $\theta_{13}$ decreases the estimated
influence of $\theta_{13}$ compared to what would have been
obtained for a flat probability distribution.

\section{Reactor data}
\label{sec:newkamland}

We compare in section~\ref{subsec:kamlandallowed}  the allowed
oscillation regions for $\Delta m^2_{21}$, $\tan^2 \theta_{12}$,
and $\sin^2 \theta_{13}$ that are determined from the first
KamLAND results~\cite{kamlandfirstpaper}, together with the CHOOZ
data~\cite{chooz},  with the oscillation regions determined by
including the recently released new KamLAND
data~\cite{newkamland}.   This Before-After comparison is
illustrated in the two right-hand  panels of
figure~\ref{fig:solar_kamland}. We also show in
section~\ref{subsec:kamlandallowed} that the new KamLAND data more
strongly disfavor a proposed~\cite{nsi} non-standard description
of solar neutrino oscillations.

We describe in section~\ref{subsec:fivekamland} six different
sets of assumptions that were used in analyzing the KamLAND data.
We then discuss in section~\ref{subsec:kamlandtechnical} some
technical aspects of the analysis of the second release of KamLAND
data.

All of the neutrino properties and the solar neutrino fluxes that
are determined in section~\ref{sec:newglobal} are robust with
respect to the six different analysis approaches described in
section~\ref{subsec:fivekamland}.

\subsection{Allowed regions: KamLAND  reactor data}
\label{subsec:kamlandallowed}

In this subsection, we discuss briefly in
section~\ref{subsubsec:kamlandspectral} the implications of the
spectral distortion observed recently by the KamLAND
collaboration, and then in section~\ref{subsubsec:allowedkamland}
present the best-fit values for neutrino parameters and their
uncertainties, as well as the allowed contours obtained using the
new KamLAND data. We show in section~\ref{subsubsec:nonstandard}
that the new KamLAND data more strongly disfavor a previously
proposed description of solar neutrino oscillations in terms of
non-standard interactions.

\subsubsection{Spectral distortion}
\label{subsubsec:kamlandspectral}

 The new KamLAND
data~\cite{newkamland} confirm the expected deficit of $\overline
\nu_e$ due to oscillations with parameters in the LMA region. More
importantly, the new data show  the expected distortion of the
energy spectrum. In their new paper~\cite{newkamland} , the
KamLAND collaboration report a goodness-of-fit test for a scaled
no-oscillation energy spectrum with the normalization fitted to
the data.  They find a goodness-of-fit of only 0.1\%. We confirm
that the hypothesis of an undistorted scaled spectrum can fit the
data with less than 0.2\% probability.

As a consequence, the 3$\sigma$ region from the After KamLAND-only
analysis shown in the lower right hand panel of
figure~\ref{fig:solar_kamland} does not extend to mass values
larger than $\Delta m^2_{21}=2\times 10^{-4}$ eV$^2$.  For the
now-excluded large $\Delta m^2_{21}$ values, the  predicted
spectral distortions are too small to fit the KamLAND data.

 \FIGURE[!t]{
\centerline{\psfig{figure=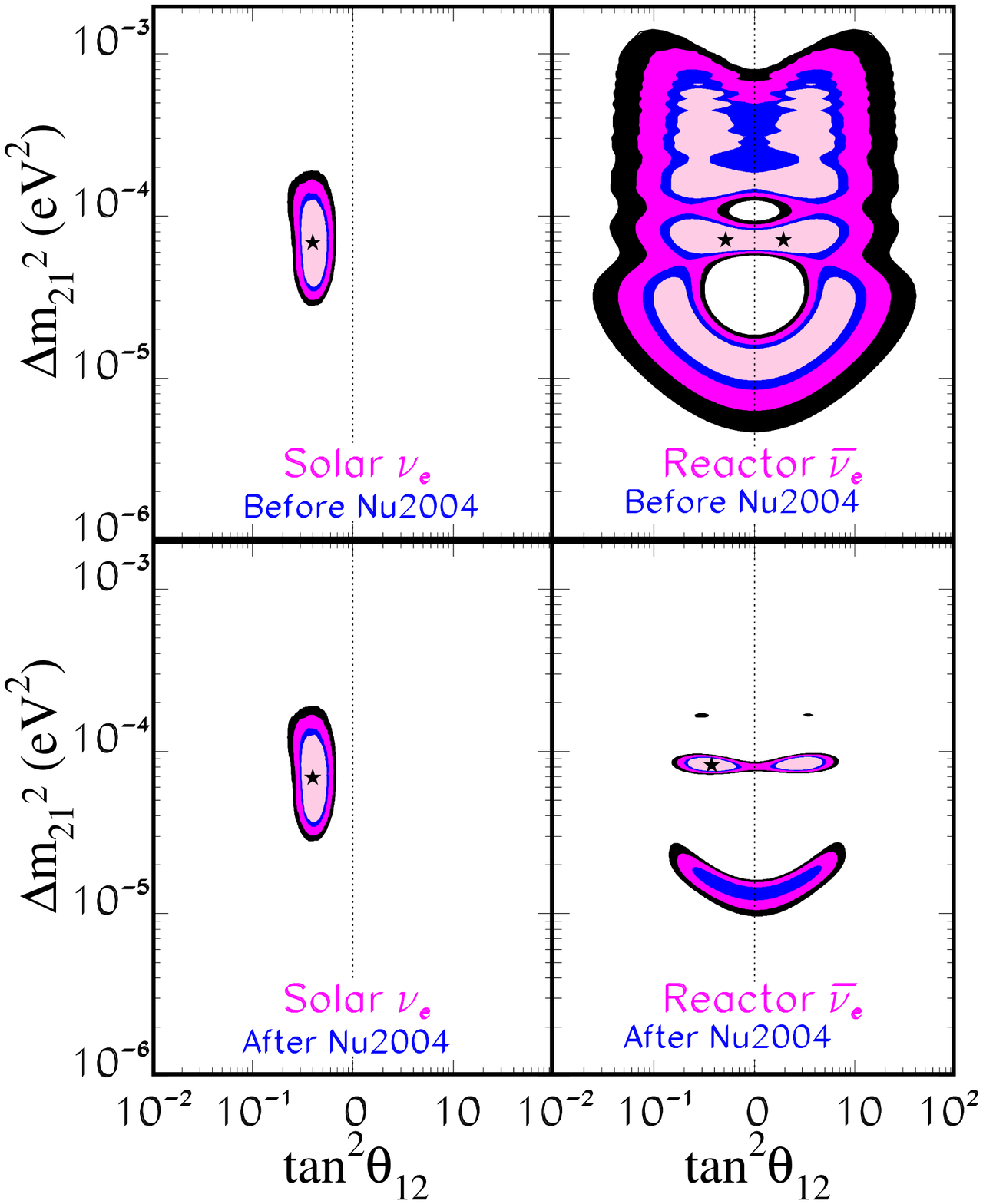,width=3.5in}}
\caption{{\bf Allowed oscillation parameters: Solar vs KamLAND.}
The two left panels show the 90\%, 95\%, 99\%, and 3$\sigma$
allowed regions for oscillation parameters that are obtained by a
global fit of all the available solar
data~\cite{chlorine,sage02,gallex,superk,snoccnc,snodaynight,gno,snosalt}.
The two right panels show the 90\%, 95\%, 99\%, and 3$\sigma$
allowed regions for oscillation parameters that are obtained by a
global fit of all the reactor data from KamLAND and
CHOOZ~\cite{kamlandfirstpaper,chooz}. The two upper (lower) panels
correspond to the analysis of all data available before (after)
the Neutrino 2004 conference, June 14-19, 2004 (Paris). The new
KamLAND data~\cite{newkamland} are sufficiently precise that
matter effects discernibly break the degeneracy between the two
mirror vacuum solutions in the lower right panel.
\label{fig:solar_kamland}}}

\subsubsection{KamLAND-only: best-fit values, uncertainties, and
allowed  regions} \label{subsubsec:allowedkamland}

 Figure~\ref{fig:solar_kamland} compares the allowed regions for
anti-neutrinos as determined by the KamLAND  reactor experiment
before Neutrino~2004 (upper right panel) with the allowed regions
after Neutrino~2004 (lower right panel). The two panels in the
right column of figure~\ref{fig:solar_kamland} represent the
Before-After summary of the effect of the new KamLAND data
released at Neutrino~2004.

Before Neutrino~2004, the best-fit solutions for the reactor data
were $\Delta m^2_{21} = 7.1 \times 10^{-5}\, {\rm eV^2}$ and
$\tan^2 \theta_{12} = 0.52\, \& \,  1.9\, $ (see
ref.~\cite{postkamland}). Within the statistical precision of the
first KamLAND data,   matter effects were too small to provide a
meaningful discrimination between the two octants for
$\theta_{12}$.

After Neutrino~2004, the best-fit values ($\chi^2 = 12.5$) is shown in the lower
panel of figure~\ref{fig:solar_kamland} and is:
\begin{equation}
 \Delta m^2_{21} = 8.3^{+0.40}_{-0.30}(^{+{1.2}}_{{-1.0}})
 \times 10^{-5}\, {\rm eV^2}, \,\,
{\tan^2 \theta_{12}} = 0.36^{+0.10}_{-0.08} (^{+6.2}_{-0.2}) \,
({\rm reactor~data: After}).
 \label{eq:bestfitpostkamland}
\end{equation}
As pointed out in Ref.~\cite{postkamland}, matter effects,
although small, cannot be neglected in the analysis of  precise
KamLAND data~\cite{newkamland}. In the analysis of the currently
available data, matter effects break the degeneracy between the
two octants of the mixing angle and induce an extra $\Delta
\chi^2=0.2$ for what would otherwise be (in the absence of matter
effects) the mirror minimum at $\tan^2\theta_{12}=2.7$.

We describe in section~\ref{subsec:fivekamland} six different
approaches to analyzing the KamLAND data. The uncertainties shown
in eq.~(\ref{eq:bestfitpostkamland}) are $1\sigma$ ($3\sigma$)
errors for KamLAND analysis option number~3 of
section~\ref{subsec:fivekamland}.

The best-fit values and uncertainties of $\Delta m^2_{21}$ and
$\tan^2 \theta_{12}$ are essentially independent of the six
analysis options for KamLAND data. The best-fit value for $\Delta
m^2_{21}$ is the same to the numerical accuracy of
eq.~(\ref{eq:bestfitpostkamland}) for all six analysis options and
the range of the $1\sigma$ uncertainty varies by only about $\pm
0.15\sigma$. The best-fit value of $\tan^2 \theta_{12}$ varies by
about $\pm 0.2\sigma$ and the range of the $1\sigma$ uncertainty
varies by $\pm 0.1 \sigma$.

Our results are in good agreement with those obtained
   by the KamLAND collaboration.  They report a best fit point  at
   $\Delta m^2_{21}=8.3 \times 10^{-5}\, {\rm eV^2}$ and
   $\tan^2\theta_{12}=0.41$, which is  within the range
   of the best fit points obtained with our six analysis procedures and
   almost identical to our preferred best-fit point. Comparing the results of
   our binned analysis with the results of  the
   event-by-event maximum likelihood analysis of KamLAND,
   we find only two ``barely visible'' differences: (i) the lower "island" is
   allowed at a slightly lower CL in all of our binned analyses; (ii) the CL
   at which the "best-fit island" extends into maximal mixing is below
   95\% CL in Ref.~\cite{newkamland},   while in our binned analysis
   we find maximum mixing is slightly above or below 95\% CL depending on the
   particular analysis option we adopt.

The new KamLAND data~\cite{newkamland}, together with the
CHOOZ~\cite{chooz}, K2K~\cite{k2k}, and atmospheric~\cite{atm}
results, lead to the following allowed range of  $\sin^2
\theta_{13}$,
\begin{equation}
\sin^2\theta _{13} = 0.005^{+0.011}_{-0.005} (^{+0.045}_{-0.005} )
\, .
 \label{eq:sin13kamland}
\end{equation}
For the six different analysis options discussed in
section~\ref{subsec:fivekamland}, the best-fit value of $\sin^2
\theta_{13}$ varies by less than  $\pm 0.1 \sigma$ and the range
of the $1\sigma$ uncertainty varies by $\pm 0.1 \sigma$.  Note,
however, that the best-fit value of $\sin^2\theta _{13}$ is not
significantly different from zero.

\subsubsection{Non-standard interaction}
\label{subsubsec:nonstandard}

 Non-standard flavor-changing
neutrino-matter interactions could potentially play a profound
role in solar neutrino oscillations, even if the non-standard
interactions are much weaker than standard weak interactions. In
ref.~\cite{nsi}, Friedland, Lunardini, and Pe\~na-Garay proposed a
non-standard description of solar neutrino and reactor
oscillations that expanded the allowed regions for neutrino
oscillation parameters beyond what was allowed by standard
interactions. The preferred oscillation parameters for this
non-standard interaction are:
\begin{equation}
\Delta m^2_{21} = 1.5 \times 10^{-5}\, {\rm eV^2}, \,\, \tan^2
\theta_{12} = 0.39\ ({\rm non-standard~interactions:
ref.~\cite{nsi}}).
 \label{eq:bestnsi}
\end{equation}

After the new KamLAND measurements, this solution is  disfavored
at the 91\% CL for 2 dof.  This is a significant improvement over
the first KamLAND results, which disfavored at 78\% CL the
non-standard solution of eq.~(\ref{eq:bestnsi}).

\subsection{Six methods of analyzing KamLAND data}
\label{subsec:fivekamland}

We have analyzed  the new KamLAND data with six different
approaches. We first enumerate the six  sets of assumptions that
were used and then comment on the differences between the various
assumptions. We provide additional technical details in the
following subsection, section~\ref{subsec:kamlandtechnical}.

\begin{enumerate}
\item{Poisson statistics, our normalization, 13 energy bins}

\item{Poisson statistics, KamLAND normalization, 13 energy bins}

\item{Poisson statistics, our normalization, 9+1 energy bins}

\item{Poisson statistics, KamLAND normalization, 9+1 energy bins}

\item{Gaussian statistics,our normalization,9+1 energy bins}

\item{Gaussian statistics, KamLAND normalization, 9+1 energy bins}
\end{enumerate}

We prefer to combine the four highest energy bins of the KamLAND,
which have only 6 events in total, in order to reduce the
fluctuations and to make the analysis more stable. In order to
verify that this additional binning does not affect the final
results, we performed separate analyses with the full 13 published
KamLAND energy bins (1 and 2 above) and with 9 + 1 energy bins (3,
4, 5, and 6). As we shall see in
section~\ref{subsec:kamlandtechnical}, our best-fit normalization
for the number of observed events agrees with the best-fit
normalization of KamLAND but the two normalizations differ
slightly (by less than $1\sigma)$. We have therefore performed
analyses with our normalization (1, 3, and 5 above) and separately
with the KamLAND normalization (2, 4, and 6).  Finally, we have
compared, with the same energy binning and normalization, the
results obtained with Poisson statistics (item 3) with the results
obtained with Gaussian statistics (item 5).

Of the six possibilities listed above, we prefer number 3. This
option relies totally on our own calculations,  so it is an
independent check of the calculations of the KamLAND
collaboration~\cite{newkamland}. Moreover, option 3 minimizes the
effects of fluctuations due to low statistics bins.

Fortunately, we shall show in section~\ref{sec:newglobal} that the
globally-inferred results for neutrino parameters and solar
neutrino fluxes are essentially independent of which one of the
six options we choose. We have already seen in
section~\ref{subsec:kamlandallowed} that all six of the analysis
options yield consistent results to an accuracy of much better
than $1\sigma$.

\subsection{Some technical details: reactor anti-neutrino analysis}
\label{subsec:kamlandtechnical}

The present analysis is based on data taken from 9 March, 2002
through 11 January, 2004 \cite{newkamland}.  We take account of
corrections due to, among other things, the spallation cut and the
detection efficiency of the tagged signal of electron
antineutrinos (see ref.~\cite{newkamland}). We assume a time
independent correction due to maintenance and  bad runs and
normalize our results after fiducial cuts to the KamLAND total
exposure of 766.3 ton$\cdot$year.

We included in our calculations the time dependences due to the
turn on/off of the different reactors in Japan. We have tracked
the power of Japanese reactors on \hbox{$\rm
http://www.fepc-atomic.jp/public_info/unten/index.html$}; the web
page is owned by The Federation of Electric Power Companies of
Japan. At present, this Web page tabulates the operational days of
the 52 Japanese reactors up to June 2003. Furthermore, we have
also been tracking the reactor power on a weekly basis since April
2003. This information allowed us to account for the time
variations of the power-averaged antineutrino baseline, which can
be as large as 20\% (in good agreement with the results presented
by KamLAND~\cite{newkamland}). Other time dependences like
variations of the reactor composition could not be tracked, but
have been shown to be small~\cite{Murayama:2000iq}. We use the
time averaged fuel compositions $^{235}$U: $^{238}$U: $^{239}$Pu:
$^{241}$Pu = 0.568: 0.078 : 0.297: 0.057. Non-Japanese reactors
contribute to the KamLAND signal less than 3\% and their flux
contribution is assumed time independent.

With all this information, we find that, in the absence of
oscillations, the expected number of antineutrino  events above
2.6 MeV energy threshold is 381 which is in good agreement with
the KamLAND estimate of $365.2\pm 23.7({\rm syst})$. All of the
solar neutrino parameters we infer from a global solution of the
solar plus reactor data are, to high accuracy (much better than
$1\sigma$) independent of which normalization we adopt (see items
1 and 2, items 3 and 4, and items 5 and 6 of section~\ref{subsec:fivekamland} and
the discussion of the results all six analysis approaches in
section~\ref{sec:newglobal}).

We analyze the KamLAND energy spectrum by making a $\chi^2$ fit to
their  binned energy spectrum. The KamLAND spectrum contains a
total of 13 energy bins above 2.6 MeV, with only 6 events in the
four highest energy bins. In order to reduce the fluctuations
associated with the small number  of events in these four bins, we
combine for options 3, 4, 5, and 6 of
section~\ref{subsec:fivekamland} the data of these high energy
events into a single bin with $E>6$ MeV (containing 6 events). For
this 10 bin analysis,  we compute the results assuming that the
binned data is Poisson distributed,
\begin{equation}
\chi^2_{\rm KamLAND}~=~{\rm min}_{\alpha}\sum_{i=1}^{10} \left[
2(\alpha R^i_{\rm th}- R^i_{\rm exp})+ 2 R^i_{\rm exp}
\ln\left(\frac{R^i_{\rm exp}}{\alpha R^i_{\rm th}} \right)\right]
+ \frac{(\alpha-1)^2}{\sigma_{\rm sys}^2} \;,
\label{eq:chi2kamland10p}
\end{equation}
or that the binned data is Gaussian distributed,
\begin{equation}
\chi^2_{\rm KamLAND}~=~{\rm min}_{\alpha}\sum_{i=1}^{10}
\frac{(\alpha \, R_{\rm th}^i -R_{\rm exp}^i)^2}{ \sigma^2_{\rm
stat,i}}+ \frac{(\alpha -1)^2}{\sigma^2_{\rm syst}}
\label{eq:chi2kamland10g}
\end{equation}
with $\sigma^2_{\rm stat,i}=R_{\rm exp}^i$. We also use a $\chi^2$
completely analogous to eq.~(\ref{eq:chi2kamland10p}) when analyzing
all 13 energy bins (options 1 and 2 of
section~\ref{subsec:fivekamland}).

In eqs.~(\ref{eq:chi2kamland10p}) and~(\ref{eq:chi2kamland10g}),
$\alpha$ is an absolute normalization constant and $\sigma_{\rm
syst}=6.5\%$ is the total systematic uncertainty from several
theoretical and experimental sources (see table I of
ref.~\cite{newkamland}). In our binned analysis we have neglected
the shape distortion errors. Using the presently available
information we have been able to compute the shape distortion
errors due to the uncertainties in the energy scale and reactor
$\overline{\nu}_e$ spectra and found them to be $<0.8\%$  and
$<0.5\%$ respectively in any of the bins. We include in $R_{\rm
th}^i$ the expected number of events in the  presence of
oscillations, including the backgrounds from accidental
coincidences ($2.69 \pm 0.02$ events) and spallation sources ($4.8
\pm 0.9$ events). The accidental background contributes to the
event rate in the first bin  while the spallation background is
distributed among all the energy bins and peaks at $E\sim 5.6$
MeV.

Our statistical analysis is different from the one of the KamLAND
collaboration.  First,  we include the effect of $\theta_{13}$ as
described in Sec.~\ref{sec:theta13}.   Second, KamLAND
collaboration performs an unbinned maximum likelihood fit.  Such
an event-by-event likelihood
    analysis provides a more powerful tool to extract information from
    the data (see for instance Ref.\cite{thomas}). At the
moment,  only the KamLAND collaboration can perform an
eventy-by-event
    maximum likelihood fit  since to do so requires
    knowing the antineutrino energy (and time) for each event, which is not
    publicly available.

 In our previous studies~\cite{postkamland,roadmap,chi2a,chi2b},
 we used a calculational grid of
 80 points per decade of $\Delta m_{21}^2$ and  80 points per decade of $\tan^2 \theta_{12}$.
 The previous grid is not sufficiently dense to take full account of the accuracy in the
 currently available  neutrino data. Hence we are now using throughout the present paper
 a grid of 180 points
 for each decade of $\Delta m_{21}^2$,   180 points for each decade of $\tan^2 \theta_{12}$,
 and a step size of 0.00125 for $\sin^2 \theta_{13}$.

As mentioned before, matter effects cannot be neglected in the
present analysis of KamLAND data. They are most important in the
lowest {\sl island} and slightly favor the {\sl light} versus the
{\sl dark} side of the mixing angle. To estimate the size of
matter effects in the present analysis we define, $F({\rm
matter~vs~vacuum})$, as the fractional difference in the event
rate for the KamLAND detector calculated with and without
including matter effects in the Earth. We find that, within the
$1\sigma (3\sigma)$ allowed region of the analysis of the KamLAND
energy spectrum,  the maximum value of $|F({\rm
matter~vs~vacuum})|$ corresponds to $0.4$\% ($2.3$\%). The maximum
change in $\chi^2$ due to including matter effects in the
KamLAND-only analysis is 0.5 (1.4) at $1\sigma (3\sigma)$.

\section{Solar neutrino analysis}
\label{sec:newsolar}

In this section, we compare the allowed oscillation regions
determined from all previous solar neutrino experiments (chlorine,
Kamiokande, SAGE, GALLEX/GNO, Super-Kamiokande,
SNO)~\cite{sage02,gallex,gno,chlorine,kamiokande,superk,snosalt,snoccnc,snodaynight}
with the oscillation regions determined by including  the slightly
revised average gallium rate released at
Neutrino~2004~\cite{sage04} with the previously available data.

We present in section~\ref{subsec:solarallowed}  the main
scientific results of this solar-only analysis . In
section~\ref{subsec:technicalsolar}, we describe some technical
details of our analysis of the solar neutrino data.

\subsection{Allowed regions: solar neutrinos}
\label{subsec:solarallowed}

 How much have the new solar neutrino data
changed the allowed regions?  The answer is 'imperceptibly', as
the reader can easily see by comparing the upper left panel of
figure~\ref{fig:solar_kamland} with the lower right panel of
figure~\ref{fig:solar_kamland}.  We challenge even the most sharp
eyed of our colleagues to discern the difference.

Figure~\ref{fig:solar_kamland} shows the allowed regions for all
solar neutrino experiments before Neutrino~2004 (upper left panel)
with the allowed regions after Neutrino~2004 (lower left panel).
The two panels in the left column of
figure~\ref{fig:solar_kamland} represent the Before-After summary
of the effect of the new SNO data released at Neutrino~2004.

The allowed regions for solar neutrino oscillations presented in
figure~\ref{fig:solar_kamland} are somewhat larger
 than the regions obtained by other
authors~\cite{otherspostsnoII}.  The reason is that we have
allowed all of the neutrino fluxes to be free parameters subject
only to the luminosity constraint~\cite{luminosity}, which is
equivalent to energy conservation if light element fusion is the
source of the solar luminosity. Most other
groups~\cite{otherspostsnoII} incorporate in their analysis the
solar neutrino fluxes and their uncertainties that are predicted
by the standard solar model~\cite{bp04,bp00}.

One can give good arguments for either including, or not
including, the predicted solar model fluxes in the
phenomenological analysis. The sound velocities measured from
helioseismology are in excellent agreement with the standard solar
model predictions~\cite{bp00,reliablemodels} and the SNO
measurement of the total $^8$B neutrino
flux~\cite{snosalt,snoccnc} is also in agreement with the solar
model predictions. These confirmations of the solar model justify
the inclusion of the solar model predictions either as priors or
as part of the $\chi^2$ analysis.

We prefer instead to allow all of the solar neutrino fluxes
to be free parameters in order to separate cleanly the astronomy
from the neutrino physics. However, we have calculated the allowed
ranges of neutrino oscillation parameters and neutrino fluxes both
ways, including the solar model predictions and letting all the
neutrino fluxes be free parameters. Both methods yield
similar--but not identical--results for $\Delta m^2_{21}$ and
$\tan^2\theta_{12}$, although the method with free fluxes and the
luminosity constraint yields a more accurate determination of the
$p-p$ solar neutrino flux~\cite{roadmap}.

\subsection{Some technical details: solar neutrino analysis}
\label{subsec:technicalsolar}

Details of our solar neutrino analyses have been described in
previous papers~\cite{postkamland,cnopaper,roadmap}. The solar
neutrino data we use are described in
table~\ref{tab:experimental}.
 Solar data includes
the Gallium (1 data point) and Chlorine (1 data point)
radiochemical rates, the Super-Kamiokande zenith spectrum (44
bins), and SNO data previously reported for phase 1 and phase 2.
The SNO data set available so far consists of the total day-night
spectrum measured in the pure D$_2$O phase (34 data points), plus
 the total charged current (CC, 1 data point), electron scattering (ES, 1 data point), and neutral
current (NC, 1 data point) rates measured in the salt
phase~\cite{snosalt,snoccnc,snodaynight}. We use for the
radiochemical experiments the neutrino absorption cross sections
given in refs.~\cite{b8spectrumbahcall,GaCS}.

We discuss in section~\ref{subsubsec:solarstrategy} our treatment
of solar neutrino fluxes. We discuss  in
section~\ref{subsubsec:sensitivity8b} how the choice of different
available determinations of the shape of the $^8$B neutrino energy
spectrum affects  the neutrino parameters and solar neutrino
fluxes that are inferred using the existing solar neutrino and
reactor anti-neutrino data.

\subsubsection{Treatment of solar neutrino fluxes}
\label{subsubsec:solarstrategy}

Total neutrino fluxes  are not required in our analysis and we
only use the model fluxes to make dimensionless the neutrino flux
output of our analysis. We express all neutrino fluxes determined
by our phenomenological analysis of experimental data as ratios of
the measured to the predicted (by the standard solar model
BP04,~\cite{bp04}) neutrino fluxes. As a result of an obsessive
sense for precision, we have used the most up-to-date electron and
neutron densities and distributions of neutrino fluxes that are
available on http:/www.sns.ias.edu/~jnb.  We have checked,
however, that none of our conclusions are affected significantly
($<$0.5\%) by the particular choice of profiles we adopt.  We have
used solar models available at http:/www.sns.ias.edu/$\sim$jnb
from 2004, 2000, 1998, and 1995. Within the accuracy of the
parameter determinations and the published grid sizes of the
models, our inferences about neutrino parameters are unaffected by
the choice of profile.  All of the profiles are essentially
unchanged to the accuracy of interest for solar neutrino work by
the improvements in the models over time (although the grid size
has increased monotonically since 1962). We may regard the
temperature and density profiles as experimentally confirmed
because of the remarkable agreement between standard solar model
predictions and helioseismological
measurements~\cite{bp00,reliablemodels}.

All neutrino oscillation parameters and neutrino fluxes are
treated as free parameters. We obtain the allowed ranges of a
particular parameter, solar or neutrino, by marginalizing the
$\chi^2$ over all other parameters. In the $\chi^2$ analysis, the
SNO and Super-Kamiokande sectors are correlated by the theoretical
$^8$B spectrum.

\subsubsection{Sensitivity to $^8$B neutrino spectrum}
\label{subsubsec:sensitivity8b}

In our previous global studies of solar neutrino and reactor
anti-neutrinos, we have used the Ortiz~et
al.~\cite{b8spectrumortiz} central values of the $^8$B neutrino
energy spectrum and the Bahcall~et~al.
errors~\cite{b8spectrumbahcall} on the energy spectrum. And,
indeed, we used this prescription in all of the initial
calculations in this paper.

Recently, however, Winter et al.~\cite{new_8Bspectrum} have
redetermined the $^8$B neutrino spectrum from new measurements of
the $\alpha$ energy spectrum following the $\beta^+$ decay of
$^8$B. We have redone our three neutrino analyses replacing the
Ortiz el al. central values of the $^8$B neutrino spectrum by the
Winter et al. central values. We find that the Gallium and
Chlorine rates are shifted downward by less than 0.1 $\sigma$. For
$^8$B neutrinos, the charged current and neutral current rates on
${\rm D_2O}$, as well as the $\nu-e$  scattering rate,  are
shifted downward by less than $0.1\sigma$, $0.3\sigma$, and $0.1
\sigma$, respectively. The higher energy bins of the charged
current and the electron scattering spectrum can be shifted
downward by less than or equal to  5\%.  With the current state of
the solar neutrino measurements,  all of these changes are small
compared to the uncertainties  that result from a $1\sigma$ shift
in the experimental (electron) energy scale (up to 20\% at
$1\sigma$) and energy resolution errors (up to 16\% at $1\sigma$).

We have verified that the values of neutrino parameters and the
values of neutrino fluxes  derived in this paper from the
different analyses are, to the accuracy we quote the numbers,
independent of whether we use the Ortiz~et al.  $^8$B neutrino
energy spectrum or the Winter~et al. energy spectrum. We have also
verified that our results are unchanged to the quoted numerical
accuracy if we use the Winter~et al. error estimate on the  shape
of the energy spectrum or the more conservative Bahcall~et~al.
error estimate.

\section{New global solution: solar plus reactor data}
\label{sec:newglobal}

In this section, we present the results of a global analysis of
all the available solar and reactor data.  We use the data
summarized in table~\ref{tab:experimental} and the total
$\chi^2_{\rm global}$ defined by
eq.~(\ref{eq:chisquaredpowerful}). We marginalize $\chi^2_{\rm
global}$ with respect to $\theta_{13}$ by making use of the
function $\chi^2_{\rm CHOOZ+ATM + K2K}(\theta_{13})$ that is
obtained from an analysis of atmospheric~\cite{atm}, K2K
accelerator ~\cite{k2k}, and CHOOZ reactor~\cite{chooz} data (see
Appendix for details).

We begin in section~\ref{subsec:oscillationparameters} by
presenting the solar neutrino oscillation parameters that are
allowed Before and After Neutrino~2004 by the totality of existing
neutrino data. We then summarize in
section~\ref{subsec:neutrinofluxes} the allowed ranges of solar
neutrino fluxes.

\subsection{Neutrino oscillation parameters: best-fit values, uncertainties, and independence of analysis method}
\label{subsec:oscillationparameters}

\TABLE[!t]{ \caption{{\bf Global allowed regions.}
The table
presents the best-fit oscillation parameters and the 1$\sigma$
(3$\sigma$) ranges determined from solar and reactor data
available Before Neutrino~2004 (top row) and After Neutrino~2004
(bottom row). We have included the results of the  marginalized
$\chi^2$ distributions for $\theta_{13}$ derived from the analysis
of Atmospheric, K2K and CHOOZ data ~\cite{noon04}. The quantity
$\sin^2\eta$ characterizes the sterile fraction (see
ref.~\cite{postkamland,roadmap} and the discussion in
section~\ref{subsec:sterile}). The table presents the value
obtained when all of the solar neutrino fluxes are treated as free
parameters and the luminosity constraint is imposed; a slightly
stronger limit is found if the solar model predictions of the
neutrino fluxes are taken into account.
The value of $\chi^2_{\rm min} = 79.9 ({\rm before})$ and
86.0(after) Neutrino~2004. \label{tab:globalosc}}
\begin{tabular}{lcccc}
\hline\noalign{\smallskip} Analysis & $\Delta m_{21}^2 (10^{-5}
eV^2)$&$\tan^2\theta_{12}$
& $\sin^2\theta_{13}$ & $\sin^2 \eta$ \\
\noalign{\smallskip}\hline\noalign{\smallskip} Before  &
$7.3^{+0.4}_{-0.6}$ ($^{+7.7}_{-2.0}$)
&$0.41^{+0.05}_{-0.05}$ ($^{+0.22}_{-0.13}$) & $<0.019$ ($<0.054$)
& $<0.10$ ($<0.38$)  \\
\noalign{\medskip} After & $8.2^{+0.3}_{-0.3}$ ($^{+1.0}_{-0.8}$)
&$0.39^{+0.05}_{-0.04}$ ($^{+0.19}_{-0.11}$) & $< 0.015 $ ($<0.048$) & $< 0.09$  ($<0.35$) \\
\noalign{\smallskip}\hline
\end{tabular} }

Table~\ref{tab:globalosc} gives the best-fit oscillation
parameters and their uncertainties determined from a
three-neutrino global fit to all the available solar and reactor
data.  We compare results obtained with the data available Before
Neutrino~2004 (upper row) with the results obtained using also new
data that became available only during or after Neutrino~2004
(lower row).

For all of the neutrino oscillation parameters, the best-fit
values obtained After Neutrino~2004 lie well within the quoted
$1\sigma$ uncertainties.

 The uncertainty for $\Delta m^2_{21}$ represents
the only dramatic improvement between the Before and the After
analyses. The $1\sigma$ ($3\sigma$) uncertainty in $\Delta
m^2_{21}$ is reduced by a factor of 1.7 (5.1). This reduction in
the uncertainty of $\Delta m^2_{21}$ is almost entirely due to the
observation of a statistically significant spectral energy
distortion in the new KamLAND data~\cite{newkamland}. For all the
quantities shown in table~\ref{tab:globalosc} except for  $\Delta
m^2_{21}$, the After uncertainties have been reduced by $\leq
25$\% with respect to the Before uncertainties.

The neutrino parameters inferred from the global analysis are
insensitive to which of the six analysis options discussed in
section~\ref{subsec:fivekamland} we use for the  KamLAND data. The
inferred values of $\Delta m^2_{21}$ and their uncertainties are
identical for all six options to the numerical accuracy shown in
table~\ref{tab:globalosc}.  The best-fit value for $\tan^2
\theta_{12}$ ($\sin^2 \theta_{13}$) varies by $0.3\sigma$ ($<
0.1\sigma$) for the six options and the $1\sigma$ and $3\sigma$
uncertainties vary by only 10\% ($<10$\%).
\FIGURE[!t]{
\centerline{\psfig{figure=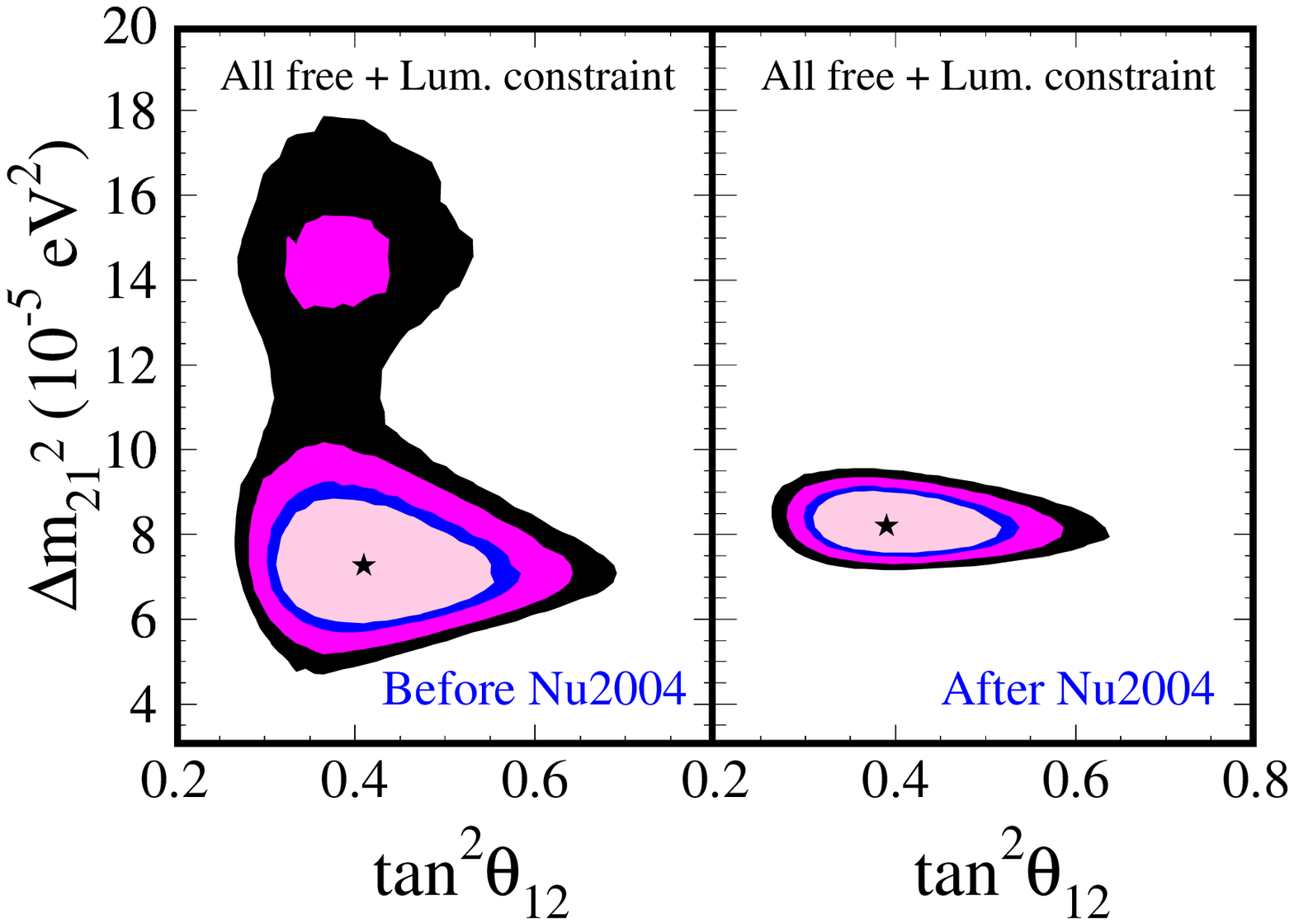,width=3.5in}}
\caption{{\bf `Before' and `After' oscillation parameters: Solar
plus reactor measurements.} The figure shows the 90\%, 95\%, 99\%,
and 3$\sigma$ allowed regions for oscillation parameters that are
obtained by a global fit of  the available
solar~\cite{chlorine,sage02,gallex,superk,snoccnc,snodaynight,gno,snosalt}
and reactor data~\cite{newkamland,kamlandfirstpaper,chooz}. Left
(Right) panel corresponds to analysis of data available before
(after) Neutrino 2004 conference. The `after' panels contain data
from either ref.~\cite{newkamland} (new KamLAND data) or
ref.~\cite{sage04} (new gallium data).\label{fig:SplusK}}}

Figure~\ref{fig:SplusK} compares the allowed regions of solar
neutrino oscillation parameters that were permitted before
Neutrino~2004 with the allowed regions that are permitted after
Neutrino~2004.  The  upper ($\sim 1.4\times10^{-4}\, {\rm eV^2}$)
allowed region and the connecting region between the upper and
lower ($\sim 7\times10^{-5}\, {\rm eV^2}$) allowed regions, all of
which is permitted by the data existing before Neutrino~2004, is
disfavored at more than $3\sigma$ after the  data presented at
Neutrino~2004 are included. The elimination of the higher $\Delta
m^2_{21}$ regions is  due to the greater precision of the new
KamLAND data~\cite{newkamland}.

Maximal mixing is strongly excluded both before and after
Neutrino~2004,

\begin{equation}
\tan^2 \theta_{12} < 1.0 ~{\rm at} ~5.5 \sigma(5.8 \sigma)
~~~~{\rm Before~(After)}. \label{eq:maximalmixing}
\end{equation}

\subsection{Sterile neutrinos}
\label{subsec:sterile}

We  consider in this subsection the constraints on the admixture
of sterile neutrinos. We concentrate on scenarios  in which all
$\Delta m^2$'s but one  are large enough to be averaged out in the
solar and reactor neutrino experimental setups~\cite{four}
\footnote{\widowpenalty=200 The effects of sterile neutrino admixtures
with two different mass scales have been discussed for solar neutrinos
in Ref.~\cite{smirnovsterile}.}. The 4-$\nu$ models invoked to explain
the LSND signal~\cite{lsnd} together with the solar and atmospheric
neutrino data are an example of the kind of model we consider here.

There are strong constraints of $\nu_e$ admixtures with large
$\Delta m^2$'s from the lack of observation of oscillations of
reactor $\overline{\nu}_e$ at short
distances~\cite{chooz,bugey,paloverde}. These constraints imply
that in the scenarios considered here solar and reactor
oscillations can be understood as a single wavelength oscillation
of $\nu_e$ into a state that is a linear combination of active
($\nu_a$) and sterile ($\nu_s$) neutrino states, $\nu_e \to
\cos\eta \, \nu_a \,+\, \sin\eta \,\nu_s$ where $\eta$ is the
parameter that describes the active-sterile admixture. The sterile
contribution to the solar neutrino fluxes can also be
parameterized in terms of $\sin^2\eta$ or, alternatively, in terms
of a derived parameter $f_{\rm B,\,sterile}$, the sterile fraction
of the $^8$B neutrino flux (see discussion in
ref~\cite{postkamland}).

The $1(3)\sigma$ allowed range for the active-sterile admixture is
\begin{equation}
\sin^2\eta\leq 0.09 (0.35) ~{\rm at}~ 1(3)\sigma \label{eq:etacurrent}
\end{equation}
 for our global analysis of the existing solar plus reactor data.
 The fundamental parameter describing the sterile fraction is
$\sin^2\eta$ (see  ref~\cite{postkamland}). However, it is
convenient to think in terms of a sterile fraction of the flux,
$f_{\rm B,\, sterile}$ , which is potentially observable in the
Super-Kamiokande and SNO experiments. This range corresponds to
\begin{equation}
f_{\rm B,\, sterile}~=~0.0^{+0.06}_{-0.00} (^{+0.25}_{-0.00})~~({\rm
solar + reactor})\,. \label{eq:fbsterileK}
\end{equation}

The limits on sterile neutrinos are not changed significantly by
the additional data released at Neutrino~2004 (see the last column
of Table~\ref{tab:globalosc}).

\subsection{Solar neutrino fluxes}
\label{subsec:neutrinofluxes}

Table~\ref{tab:globalfluxes} presents the best-fit solar neutrino
fluxes and their uncertainties that are obtained from a global fit
to all the solar and reactor data.  The fluxes are normalized to
the predicted fluxes in the standard solar model~\cite{bp04}.

The most remarkable fact about table~\ref{tab:globalfluxes} is
that the inferred ranges of the solar neutrino fluxes are very
robust. The fluxes and their uncertainties are practically
unchanged by the data released at Neutrino~2004.  The p-p and
$^8$B solar neutrino fluxes are determined ($1\sigma$) to 2\% and
5\%, respectively, and the $^7$Be and the CNO fluxes (see below)
are very poorly determined.

\TABLE[!t]{ \caption{{\bf Solar neutrino fluxes.}
The table
presents the best-fit solar neutrino fluxes normalized to the BP04
solar model fluxes~\cite{bp04}. The 1$\sigma$ (3$\sigma$)
uncertainty ranges are determined from all the solar and reactor
data available Before (upper row) and After (lower row) the
Neutrino~2004 conference. The last entry entry in the table is the
fraction of the total solar luminosity that is associated with CNO
nuclear fusion reactions. The fit to existing solar neutrino
experiments prefers a CNO luminosity that is less than 0.1\%, see
ref.~\cite{cnopaper}. } \label{tab:globalfluxes}
\begin{tabular}{lcccc}
\hline\noalign{\smallskip} Analysis & pp & $^7$Be & $^8$B& ${\rm L}_{\rm CNO}$ \% \\
\noalign{\smallskip}\hline\noalign{\smallskip} Before Nu2004 &
$1.02^{+0.02}_{-0.02}$ ($^{+0.06}_{-0.06}$) &
$0.91^{+0.25}_{-0.62}$ ($^{+0.79}_{-0.91}$) &
$0.88^{+0.04}_{-0.04}$ ($^{+0.11}_{-0.12}$) &
$0.0^{+2.8}_{-0.0}$ ($^{+7.6}_{-0.0}$)  \\
\noalign{\medskip} After Nu2004 & $1.01^{+0.02}_{-0.02}$
($^{+0.06}_{-0.06}$) & $1.03^{+0.24}_{-1.03}$ ($^{+0.77}_{-1.03}$)
& $0.87^{+0.04}_{-0.04}$ ($^{+0.09}_{-0.11}$) &
$0.0^{+2.7}_{-0.0}$ ($^{+7.3}_{-0.0}$)  \\
\noalign{\smallskip}\hline
\end{tabular} }

All of the estimates of the solar neutrino fluxes and their
uncertainties are insensitive to which of the six KamLAND
analysis options (see section~\ref{subsec:fivekamland}) we employ.
The best-fit $p-p$ neutrino flux and the $1\sigma$ and $3\sigma$
$p-p$ flux uncertainties are the same for all six options to the
numerical accuracy of table~\ref{tab:globalfluxes}.  The best-fit
$^8$B neutrino flux varies by only $0.25\sigma$, the $1\sigma$
uncertainty is the same in all cases, and the $3\sigma$ range
varies by 5\% for all six options.  The range of solutions found
for the $^7$Be solar neutrino flux using the six different
analysis options is negligible compared to the  uncertainties that
arise from the experimental data and that are shown in
table~\ref{tab:globalfluxes}.

The global analysis yields the following values for the CNO fluxes
using data available before Neutrino~2004:
\begin{equation}
^{13}{\rm N} = 0.0^{+4.3}_{-0.0}; ~~^{15}{\rm O} =
0.0^{+2.4}_{-0.0}; ~~^{17}{\rm F} = 0^{+2.2}_{-0.0}  ~~~~{\rm
Before}. \label{eq:beforecnofluxes}
\end{equation}

Using all the data available after Neutrino~2004, the CNO fluxes
are:
\begin{equation}
^{13}{\rm N} = 0.0^{+4.1}_{-0.0}; ~~^{15}{\rm O} =
0.0^{+2.3}_{-0.0}; ~~^{17}{\rm F} = 0^{+2.1}_{-0.0}  ~~~~{\rm
After}. \label{eq:aftercnofluxes}
\end{equation}

The global fit to all the data yields  a best-fit CNO fractional
contribution to the solar luminosity that is less than $0.1$\%,
much less than the range, 0.8\% to 1.6\%, implied~\cite{bp04} by
standard solar model calculations. However, the constraint shown
in table~~\ref{tab:globalfluxes} is consistent with the BP04 solar
model prediction at $1\sigma$.

\subsection{How sensitive are the global results to the precise gallium
event rate?} \label{subsec:galliumsensitivity}

\TABLE[h!t]{\caption{Analysis of solar and reactor data assuming
different  values of the  event rate in gallium solar neutrino
experiments. The table presents results for three different
assumed gallium solar neutrino event rates; all the other solar
and reactor data available prior to Neutrino~2004 were used
unchanged in the analyses. The three cases considered are
 a low gallium rate (average of GNO and SAGE for the period
1998-2003), the average gallium rate (GALLEX/GNO and SAGE
1991-2003), and a high gallium rate (GALLEX and SAGE 1991-1997).
The solar neutrino fluxes are treated as free parameters,
constrained by energy conservation (the luminosity constraint),
and are normalized to the BP04 predicted
fluxes.\label{tab:galliumsensitivity}}
\begin{tabular}{@{\extracolsep{11pt}}lccc}
\hline \noalign{\smallskip} Gallium Rate (SNU) & $63.3 \pm 3.6$ & $68.1 \pm 3.75$ & $77.8 \pm 5.0$ \\
\hline \noalign{\smallskip}\noalign{\smallskip} $\Delta m^2_{21}
(10^{-5} eV^2)$ & $8.2^{+0.3}_{-0.3}$ ($^{+1.0}_{-0.8}$)&
 $8.2^{+0.3}_{-0.3}$ ($^{+1.0}_{-0.8}$)& $8.2^{+0.3}_{-0.3}$ ($^{+1.0}_{-0.8}$)\\
\noalign{\smallskip}\noalign{\smallskip} $\tan^2\theta_{12}$ &
$0.39^{+0.05}_{-0.04}$ ($^{+0.19}_{-0.11}$)
& $0.39^{+0.05}_{-0.04}$ ($^{+0.19}_{-0.11}$) & $0.38^{+0.05}_{-0.05}$ ($^{+0.21}_{-0.11}$) \\
\noalign{\smallskip}\noalign{\smallskip} $p-p$ &
$1.03^{+0.02}_{-0.02}$ ($^{+0.05}_{-0.07}$) &
$1.01^{+0.02}_{-0.02}$ ($^{+0.06}_{-0.06}$) & $0.99^{+0.02}_{-0.02}$ ($^{+0.07}_{-0.06}$) \\
\noalign{\smallskip}\noalign{\smallskip}$^8$B
&$0.87^{+0.04}_{-0.04}$($^{+0.09}_{-0.11}$)
&$0.87^{+0.04}_{-0.04}$($^{+0.09}_{-0.11}$) &$0.88^{+0.04}_{-0.04}$($^{+0.09}_{-0.12}$)\\
\noalign{\smallskip}\noalign{\smallskip}$^7$Be&$0.25^{+0.85}_{-0.25}$($^{+1.37}_{-0.25}$)
&$1.03^{+0.24}_{-1.03}$ ($^{+0.77}_{-1.03}$) & $1.29^{+0.26}_{-0.57}$ ($^{+0.74}_{-1.29}$) \\
\noalign{\smallskip} \hline
\end{tabular}
}

The average rate of the two gallium experiments, GALLEX/GNO and
SAGE, was slightly higher in the earlier period of data taking,
1991-1997, than it was in the later period,
1998-2003~\cite{gavrin,cattadori}. In the earlier period, the average rate
of the two experiments was \hbox{($77.8 \pm 5.0$) SNU}, while in
the latter period, the average rate was \hbox{($63.3 \pm 3.6$)
SNU}~\cite{gavrin}. The grand average for the period in question,
1991-2003, is \hbox{($68.1 \pm 3.75$) SNU}.

V. Gavrin~\cite{gavrin} has raised the possibility that the
difference between the observed rates in the earlier period and in
the latter period could be due to a time-dependence of the
lower-energy $p-p$ and $^7$Be fluxes.  In this phenomenological
interpretation, the hypothesized variation could not be observed
in other solar neutrino experiments because of a lack of
sensitivity to the $p-p$ or $^7$Be solar neutrinos or because of
the long time scale and small amplitude of the assumed time
dependence. Only further long-term measurements with low energy
solar neutrino detectors can settle experimentally this question.
We address here instead a related issue.

We investigate in this subsection the extent to which the precise
value of the gallium rate influences the inferred neutrino
properties and solar neutrino fluxes that are obtained by a global
solution to all the available solar and reactor data.

Table~\ref{tab:galliumsensitivity} presents the results of three
different global analyses corresponding to the three different
assumed gallium event rates: \hbox{($77.8 \pm 5.0$) SNU}
(1991-1997), \hbox{($68.1 \pm 3.75$) SNU} (1991-2003), and
\hbox{($63.3 \pm 3.6$) SNU} (1998-2003). All the other solar
neutrino and reactor anti-neutrino data available prior to
Neutrino~2004 were used unchanged in the analyses, which were
performed as described in section~\ref{subsec:technicalsolar}.

The results obtained from the global analyses are remarkably
robust with respect to the assumed gallium event rate. Within the
range considered from 63 SNU to 78 SNU,   the inferred values for
$\Delta m^2_{21}$ and $\tan^2 \theta_{12}$ are essentially the
same for all three choices of the gallium rate. The three
best-constrained solar neutrino fluxes, $p-p$, $^8$B, and $^7$Be,
also change very little (changes less than $1\sigma$)as the
gallium rate is varied within the allowed range.

\section{Predicted rates for $^7$Be, $p-p$, and $pep$ experiments}
\label{sec:predictedrates}

In this section, we summarize the predicted rates for future
$\nu-e$ scattering experiments with $^7$Be, $p-p$, and $pep$ solar
neutrinos.

Whether or not the $^7$Be solar neutrino flux is treated as a free
parameter affects strongly the uncertainty, and the best estimate,
of the predicted rate in a $^7$Be $\nu-e$ scattering experiment.
If we perform the global analysis of solar plus reactor
experiments assuming that the BP004 calculations for the solar
neutrino fluxes and their uncertainties are valid then the
predicted rate in a $^7$Be $\nu-e$ scattering
experiment~\cite{borexino} is, with $1\sigma$ ($3\sigma$)
uncertainties:

\begin{equation}
\left[{\rm ^7Be}\right]_{\rm \nu-e}~=~ 0.665\pm 0.015\,
(^{+0.045}_{-0.040})\,\, [{\rm BP04~prediction}].
 \label{eq:be7esprediction}
\end{equation}
Here $\left[{\rm ^7Be}\right]_{\rm \nu-e}$ is the $\nu-e$
scattering rate in units of the rate that would be expected if the
BP04 solar model were exactly correct and  neutrinos did not
oscillate. Thus, if we use the solar model calculation of the
$^7$Be neutrino flux and its uncertainty, then the predicted event
rate for the $^7$Be rate experiment has a precision of $\pm 3$\%.
If instead we treat all of the neutrino fluxes as free parameters,
then the uncertainties are much larger:
\begin{equation}
\left[{\rm ^7Be}\right]_{\rm \nu-e}~=~ 0.63^{+0.13}_{-0.24}
(^{+0.44}_{-0.57})\,\, {\rm [all~fluxes~free +
luminosity~constraint]}\, .
 \label{eq:be7allfreeesprediction}
\end{equation}
The result for $\nu-e$ scattering that is given in
eq.~(\ref{eq:be7esprediction}) and
eq.~(\ref{eq:be7allfreeesprediction}) includes neutrinos from all
the solar neutrino fluxes that produce recoil electrons with
energies in the  range  0.25-0.8 MeV. Most of the recoil electrons
in the selected energy range are produced by $^7$Be neutrinos,
although there are small contributions from CNO, $pep$, and $p-p$
neutrinos (whose fluxes we estimate by taking the values from the
BP04 solar model~\cite{bp04}). The best-estimates given in
eq.~(\ref{eq:be7esprediction}) and
eq.~(\ref{eq:be7allfreeesprediction}) would be changed by much
less than $1\sigma$, to 0.664 and 0.68, respectively, if we
included only events caused by $^7$Be neutrinos.

The predictions given in eq.~(\ref{eq:be7esprediction}) and
eq.~(\ref{eq:be7allfreeesprediction}) are robust.  The results
obtained here are well within the $1\sigma$ quoted uncertainties
of the pre-Neutrino~2004 predictions (see ref.~\cite{roadmap}).

The corresponding predicted rates in a $p-p$ $\nu-e$ scattering
experiment are, with $1\sigma$ ($3\sigma$) uncertainties:
\begin{equation}
\left[{\rm p-p}\right]_{\rm \nu-e}~=~  0.707 ^{+0.011}_{-0.013}
(^{+0.041}_{-0.039})\, [{\rm BP04~prediction}]\, ,
 \label{eq:ppesprediction}
\end{equation}
and
\begin{equation} \left[{p-p}\right]_{\rm \nu-e}~=~0.716 ^{+0.014}_{-0.016} (^{+0.050}_{-0.044})
\, {\rm [all~fluxes~free + luminosity~constraint]}\, .
 \label{eq:ppallfreeesprediction}
\end{equation}

The predicted rates in a $pep$ $\nu-e$ scattering experiment are:
\begin{equation}
\left[{\rm pep}\right]_{\rm \nu-e}~=~  0.644 ^{+0.011}_{-0.013}
(^{+0.045}_{-0.037})\, [{\rm BP04~prediction}]\, ,
 \label{eq:pepesprediction}
\end{equation}
and
\begin{equation} \left[{pep}\right]_{\rm \nu-e}~=~0.652 ^{+0.015}_{-0.016} (^{+0.053}_{-0.041})\, {\rm [all~fluxes~free + luminosity~constraint]}\, .
 \label{eq:pepallfreeesprediction}
\end{equation}

The predicted rates of the $p-p$ and $pep$ solar neutrino
scattering experiments are robust~\cite{roadmap}.  The $p-p$ and
$pep$ rates are affected by less than $1\sigma$ by the choice of
whether or not we use flux predictions from the BP04 solar model
or allow the fluxes to vary as free parameters.

Since the $p-p$, $pep$, and $^7$Be solar neutrinos sample the
survival probability at different energies (as reflected by the
different values given in the predicted rates shown above), it is
important to measure $\nu-e$ scattering rates for all three
fluxes.
\section{CPT bound}
\label{sec:cpt}

In this section, we compare the regions of allowed oscillation
parameters for neutrinos and anti-neutrinos in order to constrain
the violation of CPT in the neutrino
sector~\cite{Coleman:1997xq,Coleman:1998ti,colladay,Barger:2000iv,bahcall:2002ia,Murayama:2003zw}.
Our discussion follows the pre-KamLAND analysis of
ref.~\cite{bahcall:2002ia}, in which expected results from KamLAND
were simulated.

Figure~\ref{fig:solar_kamland} shows, in the left-hand panel, the
90\%, 95\%, 99\%, and 3$\sigma$ allowed regions for oscillation
parameters that are obtained by a global solution to all the
available solar data and, in the right-hand panel, to all the
available reactor data. The juxtaposition of purely solar and
purely reactor data in the same figure allows a visual comparison
of the constraints on neutrino and on anti-neutrino oscillation
parameters. The allowed regions obtained by a global solution to
all the available solar and reactor data are shown in
figure~\ref{fig:SplusK}.

We characterize, following ref.~\cite{bahcall:2002ia}, the
sensitivity of reactor and solar neutrino experiments to CPT
violation by the quantity
\begin{equation}
\langle\Delta CPT\rangle ~=~2 \frac{|\rm R_{\nu\nu}-
R_{\bar{\nu}\bar{\nu}}|} {\left[\rm R_{\nu \nu} +
R_{\bar{\nu}\bar{\nu}}\right]} \label{eq:defndeltacpt}
\end{equation}
Here both ${\rm R_{\nu \nu}}$ and ${\rm R_{\bar{\nu}\bar{\nu}}}$
are computed for the present KamLAND experimental set up, but
using, respectively,  values for ($\Delta m^2_\nu$, $\theta_\nu$)
and for ($\Delta m^2_{\bar{\nu}}$, $\theta_{\bar{\nu}}$) within
 the allowed regions of the solar-only analysis (section~\ref{subsec:solarallowed}
 and the lower left-hand panel of figure~\ref{fig:solar_kamland})
 and of the KamLAND-only analysis
 (section~\ref{subsec:kamlandallowed} and the lower right-hand panel of figure~\ref{fig:solar_kamland})
at a given CL. Then, $\langle\Delta CPT\rangle$ is the number of
events observed in KamLAND  minus the number of events that are
expected from the solar-only analysis if neutrinos and
anti-neutrinos had exactly the same oscillation parameters,
divided by the average.

The maximum value of $\langle\Delta CPT\rangle$ is
\begin{equation}
\langle\Delta CPT\rangle  ~\leq~0.52~(1.01)\ \ \ \ {\rm at} \ \
1\sigma \ \ (3\sigma). \label{eq:DeltaCPTupper}
\end{equation}
Matter effects, which simulate CPT violation,  contribute less
than 0.01 (0.05) to $\langle\Delta CPT\rangle$ for neutrino
parameters in the 1$\sigma$ (3$\sigma$) KamLAND-only  allowed
regions (lower right-hand panel of figure~\ref{fig:solar_kamland})
and 0.08 (0.09) for neutrino parameters in the Solar-only allowed
regions (lower left-hand panel of figure~\ref{fig:solar_kamland}).

This experimental upper limit on $\langle\Delta CPT\rangle$ can be
used to test arbitrary future conjectures of CPT violation.
Following ref.~\cite{bahcall:2002ia}, we consider an effective
interaction which has been discussed by Coleman and
Glashow~\cite{Coleman:1997xq,Coleman:1998ti}, and by Colladay and
Kostelecky~\cite{colladay}, that violates both Lorentz invariance
and CPT invariance. The interaction is of the form
\begin{equation}
{\mathcal{L}}(\Delta CPT)) ~=~\bar{\nu}_{\rm
L}^\alpha{b^{\alpha\beta}_\mu}{\gamma_\mu}{\nu_{\rm L}^\beta},
 \label{eq:interaction}
\end{equation}
where $\alpha, \beta$ are flavor indices, L indicates that the
neutrinos are left-handed, and $b$ is a Hermitian matrix. We
discuss the special case with rotational invariance in which $b_0$
and the mass-squared matrix are diagonalized by the same mixing
angle. We also assume that there are only two interacting
neutrinos (or anti-neutrinos) and follow previous authors in
defining $\eta$ as the difference of the phases in the unitary
matrices that diagonalize $\Delta m^2$ and the CPT odd quantity
$\delta b$, which is the difference between the two eigenvalues of
$b_0$.

When the relative phase is $\eta =0$, the survival probabilities
of neutrino and anti-neutrinos take on an especially simple form.
In the case of reactor antineutrinos, the  oscillations occur in
vacuum to an excellent approximation (matter effects are
negligible in the range of parameters allowed by KamLAND data, see
details in ref.~\cite{postkamland}):
\begin{equation}
P_{\rm ee}~=~\cos^4\theta_{13} ~\left[1 -  \sin^22\theta_{12}
\sin^2 \left(\frac{\Delta m^2_{21} L}{4E_\nu} - \frac{\delta b
L}{2}\right)\right].\label{eq:peevacuum_mod}
\end{equation}
In the case of solar neutrinos, sensitivity to $\delta b$ in the
oscillation phase is lost because of the fast oscillations due to
the $\Delta m^2_{21}$ term. In both cases, solar and reactor neutrinos,
the mixing in vacuum is unchanged. The standard survival probability
and mixing in matter are valid with a modified ratio of
matter to vacuum effects given by
\begin{equation}
\beta~=~ \frac{2 \sqrt2 G_F \cos^2\theta_{13} n_e E_\nu}{\Delta
m^2_{21} + 2\, \delta b\, E_\nu}\, . \label{eq:defbeta_cpt}
\end{equation}

We have reanalyzed solar and reactor neutrino data with the
modified probabilities that contain an extra free parameter
$\delta b$. We obtain the $\chi^2$ distribution on $\delta b$ by
marginalizing over all other variables. The resulting
 upper bound on the violation of CPT invariance is:
\begin{equation}
\delta b < 0.6 \, (1.5)\times 10^{-20} \,{\rm GeV} , ~~\eta = 0\ \
\ \ {\rm at} \ \ 1\sigma \, (3\sigma). \label{eq:deltabeta0}
\end{equation}

\section{Summary and discussion}
\label{sec:summary}

We summarize and discuss in this section our main conclusions.

The most important feature of the  recent KamLAND
results~\cite{newkamland} is the detection of the spectral energy
distortion that was expected for  oscillations with parameters
within the Large Mixing Angle region. The distortion implies that
the $3\sigma$ region from the KamLAND-only analysis no longer
extends to masses larger than $\Delta m^2_{21}=2\times 10^{-4}$
eV$^2$.

Comparing the upper and lower right hand (KamLAND-only) panels
 of figure~\ref{fig:solar_kamland}, one can see
clearly the effect of the observed spectral distortion in
eliminating previously allowed regions with a large $\Delta
m^2_{21}$. In the now-excluded large mass regions, the predicted
spectral distortions are too small to fit the KamLAND data.

As a consequence of the observation of spectral distortion, the
largest change in the globally allowed regions of neutrino
parameters is  the elimination  of the larger $\Delta m^2_{21}$
regions in the global solution.  This constriction of the globally
allowed regions can be seen visually by comparing the left and
right hand panels of figure~\ref{fig:SplusK}.  The large $\Delta
m^2_{21}$ regions were previously allowed in the global solution
at 99\% CL.

The globally allowed range of $\Delta m^2_{21}$ is reduced by a
factor of 1.7 (5) at $ 1\sigma (3\sigma)$ compared to the
previously allowed range (see figure~\ref{fig:SplusK} and
table~\ref{tab:globalosc}). The best-fit value of $\Delta
m^2_{21}$ is increased by about 15\% by the new data released at
Neutrino~2004, which represents a 1.4$\sigma$ shift from the
previous global minimum in the before Neutrino~2004 analysis.

 The allowed ranges, as well as the best-fit values,  of $\tan^2\theta_{12}$ and
$\sin^2\theta_{13}$ are, as is shown in table~\ref{tab:globalosc},
not affected significantly by the new data release. For example,
the best-fit value of $\tan^2\theta_{12}$ is changed by only $0.4
\sigma$ and the $1\sigma$ range of uncertainty is decreased by
10\%. For  $\sin^2 \theta_{13}$, the $3\sigma$ bound is decreased
by 10\%.

The amount by which maximal mixing ($\tan \theta_{12} = 1$)is
disfavored is $5.4\sigma$ from just solar neutrino measurements
(before and after Neutrino~2004) and $5.8\sigma$ from a global
solution of all the solar and reactor data available after
Neutrino 2004.

The upper limit to the parameter $\sin^2 \eta$ that characterizes
sterile neutrinos is very little affected by the recent data
release, as is also shown  table~\ref{tab:globalosc}.  Allowing
all of the neutrino fluxes to be free parameters, the present
upper bound on $\sin^2 \eta$ is equivalent to a maximum sterile
fraction of 6\% for the $^8$B solar neutrino flux (see
section~\ref{subsec:sterile}).

We conclude that the parameters which describe solar neutrino
oscillations are robustly determined, if not yet as precise as we
would like.  This is the first major result of the present
reanalysis.

Our second major result is that the allowed ranges of the $p-p$
and $^8$B solar neutrino fluxes are  robustly determined by the
existing solar neutrino and reactor data.

Table~\ref{tab:globalfluxes} shows that both the best-fit values
and the $1\sigma$ and $3\sigma$ allowed ranges of all three of the
major solar neutrino fluxes, the $p-p$, $^7$Be, and $^8$B
neutrinos, are essentially unaffected by the data released at
Neutrino~2004.  The $\pm 2$ \% precision with which the $p-p$
neutrino flux is determined is due to the luminosity constraint as
well as the neutrino measurements.

The $^7$Be and CNO neutrino fluxes are very poorly determined
(both Before and After Neutrino~2004)(see
table~\ref{tab:globalfluxes} and eq.~(\ref{eq:aftercnofluxes})).
The upper limit to the fraction of the solar luminosity that is
associated with CNO fusion reactions remains about a factor of
six above the value predicted by the BP04 solar model (see also
table~\ref{tab:globalfluxes}).

The rate of a $^7$Be neutrino-electron scattering experiment is
rather well predicted if the BP04 solar model calculations of the
solar neutrino fluxes are assumed and is rather poorly predicted
if all the solar neutrino fluxes are allowed to be free parameters
in fitting the existing solar and reactor data. If the solar model
predictions are adopted, then the predicted rate is $0.665\pm
0.015\, (^{+0.045}_{-0.040})$ of the rate implied by the BP04
solar model in the absence of neutrino oscillations. The rate
predicted if the neutrino fluxes are treated as free parameters is
$0.63^{+0.13}_{-0.24} (^{+0.44}_{-0.57})$ of the solar model
prediction without oscillations.

The predicted rates for $p-p$ and $pep$ solar neutrino experiments
are robust and essentially independent of whether one uses fluxes
from the BP04 solar model or allows the fluxes to vary freely
subject to the luminosity constraint. If we use the BP04 fluxes,
the predicted rate of a $p-p$ experiment is $\left[{\rm
p-p}\right]_{\rm \nu-e}~=~  0.707 ^{+0.011}_{-0.013}
(^{+0.041}_{-0.039})$ and the predicted rate for a $pep$
experiment is $\left[{\rm pep}\right]_{\rm \nu-e}~=~  0.644
^{+0.011}_{-0.013} (^{+0.045}_{-0.037})$.

Since the $^7$Be, $p-p$, and $pep$ fluxes sample the survival
probability curve at different energies (see the results and
discussion in section~\ref{sec:predictedrates}), it is important
to measure the $\nu-e$ scattering of all three of these fluxes.
According to equation~(8) and figure~1 of Ref.~\cite{roadmap}, all
three of these fluxes should be in the region of neutrino
oscillation space in which vacuum neutrino oscillations dominate
over matter oscillations.

CPT violation in the neutrino sector is limited by the comparison
of the allowed neutrino oscillation regions (left hand panels of
figure~\ref{fig:solar_kamland}) with the allowed anti-neutrino
oscillation regions (right hand panels of
figure~\ref{fig:solar_kamland}). This comparison leads to an
accurate experimental limit.  This limit is presented in both a
model independent and a model dependent way in
section~\ref{sec:cpt}.

We have verified that our results are insensitive (changes much
less than $1\sigma$) to which of six approaches  we use in
analyzing the KamLAND data (section~\ref{subsec:fivekamland}),
which of the published $^8$B neutrino energy spectra we adopt
(section~\ref{subsubsec:sensitivity8b}), and the precise value of
the gallium solar neutrino event rate
(section~\ref{subsec:galliumsensitivity}).

\acknowledgments JNB and CPG acknowledge support from NSF grant
No.~PHY0070928. MCG-G is supported by  National Science Foundation
grant PHY0098527 and by Spanish Grants No FPA-2001-3031 and
CTIDIB/2002/24. We are grateful to M. Maltoni and A. McDonald for
useful discussions and comments.
\appendix
\section{Analysis details regarding $\theta_{13}$}
\label{sec:theta13}

Table~\ref{tab:globalosc} presents our limit on $\theta_{13}$. In
this section, we describe some of the details of the method used
to obtain this limit.  We shall see that the $\theta_{13}$ bound
from the atmospheric and CHOOZ data implies that three neutrino
effects are small for solar and reactor experiments. However these
three neutrino effects are not totally negligible and they
contribute to the final bound on $\theta_{13}$.

The minimum joint description of atmospheric, K2K, solar, and
reactor data requires that all the three known neutrinos take part
in the oscillation process.  The mixing parameters are encoded in
the $3 \times 3$ lepton mixing matrix which can be conveniently
parameterized in the standard form:
\begin{equation}
    U=\left(\begin{array}{ccc}
    1&0&0 \\
    0& {c_{23}} & {s_{23}} \\
    0& -{s_{23}}& {c_{23}}
    \end{array}\right)\times\left(
    \begin{array}{ccc}
    {c_{13}} & 0 & {s_{13}}e^{i {\delta}}\\
    0&1&0\\
    -{ s_{13}}e^{-i {\delta}} & 0  & {c_{13}}
    \end{array}\right)
    \times \left(\begin{array}{ccc}
    c_{21} & {s_{12}}&0\\
    -{s_{12}}& {c_{12}}&0\\
    0&0&1\end{array}\right)
    \label{eq:evol.2}
\end{equation}
where $c_{ij} \equiv \cos\theta_{ij}$ and $s_{ij} \equiv
\sin\theta_{ij}$. Note that the two Majorana phases are not
included in the expression above since they do not affect neutrino
oscillations. The angles $\theta_{ij}$ can be taken without loss
of generality to lie in the first quadrant, $\theta_{ij} \in
[0,\pi/2]$.

We  know from the analysis of solar and atmospheric oscillations
that the  mass differences satisfy:
\begin{equation} \label{eq:deltahier}
    \Delta m^2_\odot = \Delta m^2_{21} \ll
|\Delta m_{31}^2|=\Delta m^2_{\rm atm}.
\end{equation}
Under this condition, the joint three-neutrino analysis
simplifies. For the solar and KamLAND experiments, the
oscillations with the atmospheric oscillation length are
completely averaged out and the survival probability takes the
form:
\begin{equation}
    P^{3\nu}_{ee}
    =\sin^4\theta_{13}+ \cos^4\theta_{13}P^{2\nu}_{ee} \, ,
\label{eq:p3}
\end{equation}
where in the Sun $P^{2\nu}_{ee}$ is obtained with the modified
density $N_{e}\rightarrow \cos^2\theta_{13} N_e$. The analysis of
solar data constrains three of the seven parameters: $\Delta
m^2_{21}, \theta_{12}$ and $\theta_{13}$.

For atmospheric and K2K neutrinos, the solar wavelength is too
long to be relevant and the corresponding oscillating phase is
negligible. As a consequence, the atmospheric data analysis
restricts $\Delta m^2_{31}\simeq \Delta m^2_{32}$, $\theta_{23}$,
and $\theta_{13}$.  The mixing angle $\theta_{13}$ is the only
parameter common to both solar and atmospheric neutrino
oscillations; it  potentially allows for some mutual influence.
The effect of $\theta_{13}$ is to add a $\nu_\mu\rightarrow\nu_e$
contribution to the atmospheric oscillations. Finally, for the
CHOOZ experiment, the solar wavelength is unobservable if $\Delta
m^2_{21} \leq 8\times 10^{-4}~ {\rm eV}^2$. The relevant survival
probability oscillates with a wavelength determined by $\Delta
m^2_{31}$ and an amplitude determined by $\theta_{13}$.

The above considerations imply that the global analysis factorizes
as
\begin{equation}
\chi^2_{\rm global}=\chi^2_{\rm solar+Kamland}(\Delta
m^2_{21},\theta_{12},\theta_{13}) +\chi^2_{\rm
atm+K2K+CHOOZ}(\Delta m^2_{31},\theta_{23},\theta_{13})\, .
\label{eq:chiglobal}
\end{equation}
Thus the 3-$\nu$ oscillation effects in the analysis of solar and
KamLAND data are obtained from the study of:
\begin{equation}
\chi^2_{global}|_{\rm marg} (\Delta
m^2_{21},\theta_{12},\theta_{31}) =\chi^2_{\rm
solar+Kamland}(\Delta m^2_{21},\theta_{12},\theta_{13})
+\chi^2_{\rm atm+K2K+CHOOZ}|_{\rm marg}(\theta_{13})\, .
\label{eq:chiglomarg}
\end{equation}
We denote by $\chi^2_{\rm atm+K2K+CHOOZ}|_{\rm marg}$ the $\chi^2$
analysis of atmospheric, K2K and CHOOZ data after marginalization
over $\Delta m^2_{31}$ and $\theta_{23}$.

The present strong bound on $\theta_{13}$ is mostly determined by
$\chi^2_{\rm atm+K2K+CHOOZ}|_{marg}(\theta_{13})$ and arises from
the non-observation of $\overline{\nu}_e$ oscillations at CHOOZ
with an atmospheric wavelength. Here we have used the results from
our updated three-neutrino oscillation  analysis~\cite{noon04} of
atmospheric~\cite{skatmlast}, K2K~\cite{k2k}, and
CHOOZ~\cite{chooz} data  which we briefly summarize.

The basic techniques employed can be found in
ref.~\cite{3nuupdate}, but we have updated the atmospheric
neutrino analysis to include the results from the final
Super-Kamiokande phase I analysis~\cite{skatmlast}. In particular
we make use of the new three-dimensional fluxes from
Honda~\cite{honda3d} as well as the improved interaction cross
sections which agree better with the measurements performed with
near detector in K2K. We have also improved our statistical
analysis of the atmospheric data (see ref.~\cite{ournewatm} for
details).

As a result of the inclusion of all of these effects, the allowed
region for $\Delta m^2_{31}$ is shifted to lower values and the
3$\sigma$ bound on $\theta_{13}$ from $\chi^2_{\rm
atm+K2K+CHOOZ}|_{marg}(\theta_{13})$ weakened slightly  (from
$\sin^2\theta_{13}\leq 0.056$ to 0.059). Also the best-fit value
of $\sin^2\theta_{13}$ is not exactly at $\sin^2\theta_{13}=0$ but
rather at $\sin^2\theta_{13}=0.006$ (although $\sin^2\theta_{13}$
differs from 0.0  only by  $0.5\sigma$). This fact that
$\sin^2\theta_{13}$ is not exactly zero is due to the atmospheric
neutrino data,  in particular to the slight excess of sub-GeV
e-like events which are better described with a non-vanishing
value of $\theta_{13}$.

The inclusion of the KamLAND and solar data further strengthens the
$\theta_{13}$  bound  as is reflected in
Table~\ref{tab:globalosc}. Before the new KamLAND and solar
results,  the $1\sigma$ ($3\sigma$) bound on $\theta_{13}$ from
the global 3--$\nu$ analysis was $\sin^2\theta_{13}<0.019
(0.054)$. This bound is now improved by about 10\% to
$\sin^2\theta_{13}<0.016 (0.048)$.

\end{document}